\documentclass[reprint,amsmath,amssymb,aps, superscriptaddress,floatfix,prd]{revtex4-2}

\usepackage{graphicx}
\usepackage{dcolumn}
\usepackage{bm}
\usepackage{amsfonts}
\usepackage{siunitx}
\usepackage{xcolor}
\usepackage{color}
\usepackage{import}
\usepackage{placeins}
\usepackage{hyperref}

\newcommand{\haschanged}[1]{\textcolor{black}{#1}} 

\begin{document}
\title{Searching for Dark Photons with Existing Haloscope Data}

\author{Sumita Ghosh}
\email{sumita.ghosh@yale.edu}
\affiliation{Department of Applied Physics, Yale University, New Haven, Connecticut 06520, USA}
\affiliation{Wright Laboratory, Department of Physics, Yale University, New Haven, Connecticut 06520, USA}

\author{E. P. Ruddy}
\affiliation{JILA, National Institute of Standards and Technology and the University of Colorado, Boulder, Colorado 80309, USA}
\affiliation{Department of Physics, University of Colorado, Boulder, Colorado 80309, USA}

\author{M. J. Jewell}
\affiliation{Wright Laboratory, Department of Physics, Yale University, New Haven, Connecticut 06520, USA}

\author{A. F. Leder}
\affiliation{Department of Nuclear Engineering, University of California Berkeley, California 94720, USA}

\author{R.H.\ Maruyama}
\affiliation{Wright Laboratory, Department of Physics, Yale University, New Haven, Connecticut 06520, USA}

\begin{abstract}
The dark (or hidden) photon is a massive U(1) gauge boson theorized as a dark force mediator and as a dark matter candidate. Dark photons can be detected with axion cavity haloscopes by probing for a power excess caused by the dark photon's kinetic mixing with Standard Model photons. Haloscope axion exclusion limits may therefore be converted into competitive dark photon parameter limits via the calculation of a corresponding dark photon to photon coupling factor. This calculation allows for an improvement in sensitivity of around four orders of magnitude relative to other dark photon exclusions and may be attained using existing data. We present how one converts haloscope axion search limits and a summary of relevant experimental parameters from published searches. In addition, we have included the code that can be used to generate our dark photon exclusion limits for the cases described in this paper. Finally, we present limits on the kinetic mixing coefficient between dark photons and the Standard Model photons based on existing haloscope axion searches.
\end{abstract}

\maketitle

\section{\label{sec:introduction}Introduction}
Astrophysical observations suggest that $\sim$85\% of the matter in the universe is dark matter~\cite{ade2016planck}. The nature of dark matter is still unknown, but the majority of models are consistent with the following: it is massive, stable, and manifests itself primarily through its gravitational interaction with the observable universe. Theoretical  models have traditionally characterized dark matter as a single, still unidentified, particle. 

One well-motivated candidate for the elementary dark matter particle is the axion~\cite{axion_review,weinbergboson}. Cavity-based axion detectors, known as haloscopes, search for a weak, narrowband signal at a frequency corresponding to the axion's currently unknown rest mass. This signal would manifest as a slight power excess in the cavity originating from the axion's predicted coupling to electromagnetism. In addition to being sensitive to axions, haloscopes are also sensitive to another viable dark matter candidate: the massive dark photon~\cite{ackerman2009dark,darkphotonreview}. The dark photon may or may not be massive, and also has a weak coupling to electromagnetism, allowing massive dark photons to have a parameter space of mass vs coupling. Only massive dark photons can be dark matter candidates, though massless dark photons are also a potential beyond Standard Model particle. The dark photon, as a vector boson, has a polarization that can be rapidly or slowly varying, and the dark photon field can be polarized or unpolarized. \haschanged{If the field polarization varies much quicker than the detector integration time, the detector sees an effective time-averaged signal rather than any individual polarization, mitigating the issue of the unknown initial polarization.} However, if the polarization is slowly varying \haschanged{compared to the measurement, the detector will only sample single values of the potential polarizations. When combined with unknown initial polarization angle, the measurement benefits from running the detector over 24 hour periods to cover all possible polarizations \cite{caputo2021dark}.}

Here we describe a procedure to convert haloscope exclusions from axion parameter space into dark photon parameter space \haschanged{including the probabilistic distribution of dark photon polarizations in the case of a slowly-varying polarized dark photon field}. We summarize the experimental conditions necessary for this procedure, and provide the required parameters and conversion factors. In Section \ref{sec:background}, we discuss the viability of the dark photon as a dark matter candidate. In Section \ref{sec:analysis}, we describe the conversion procedure and derive the required scaling factor. In Section \ref{sec:results}, we present dark photon exclusions in the 1-30~\SI{}{\micro \electronvolt} regime using existing haloscope data. For completeness, we will review the existing body of knowledge throughout the article.

\section{\label{sec:background}Dark Photons as Dark Matter}
The dark photon is a vector boson that stems from an added U(1) symmetry, one of the simplest extensions of the Standard Model~\cite{holdom1986two}. Multiple production mechanisms such as misalignment~\cite{nelson2011dark, pospelov2008bosonic} and inflation~\cite{graham2016vector, nakai2020light} allow for dark photons in the haloscope mass range to naturally produce the required relic abundance of dark matter. However, the motivation for dark photons is not contingent on their comprising all of dark matter. The addition of one or more additional U(1) symmetries to the SM is favored by many models of string theory~\cite{arvanitaki2010string, goodsell2009naturally, cicoli2011testing} for physics beyond the Standard Model (SM). Each U(1) symmetry would have a gauge boson and fermion~\cite{brahmachari2014kinetic,redondo2009massive}, akin to the dark photon and the millicharged particle~\cite{afek2020limits,carney2021trapped}, respectively. The addition of many particles could be consistent with the existence of an entire \textit{hidden sector} ~\cite{essig2013dark,foot2015dissipative} composed of multiple dark matter particles and self-interactions. The analysis described in this paper is agnostic to whether dark matter is one particle or many until Section \ref{sec:results}, where, in order to create Figure \ref{fig:haloscope_parameter_space}, we assume that dark photons comprise all or almost all of dark matter.

The standard dark photon Lagrangian is
\begin{equation}
\small
    \label{eq:lagrangian}
     \mathcal{L} \supset -\frac{1}{4} F_1^{\mu\nu} F_{1\mu\nu} - \frac{1}{4} F_2^{\mu\nu} F_{2\mu\nu} - \frac{1}{2} \chi F_1^{\mu\nu} F_{2\mu\nu} + \frac{1}{2} m^2_{\gamma'}X^2.
\end{equation}
Here, $F_1^{\mu\nu} = \partial^\mu A^\nu - \partial^\nu A^\mu$ is the electromagnetic field tensor and $F_2^{\mu\nu} = \partial^\mu X^\nu - \partial^\nu X^\mu$ is the dark photon field tensor, where $A$ and $X$ are the photon and dark photon 4-vector fields respectively. $\chi$ is the kinetic mixing coefficient between the dark photon and the SM photon and $m_{\gamma'}$ is the mass of the dark photon.

The kinetic mixing term of Equation \ref{eq:lagrangian}, with interaction strength $\chi$, suggests that dark photons can oscillate into SM photons, in a similar way to how neutrinos can oscillate between their interaction eigenstates~\cite{jaeckel2013force}. This oscillation allows for two common methods of dark photon detection: 1) generating and then detecting the dark photon and 2) detecting a dark photon from the relic abundance in the universe. The former uses a ``light shining through walls" (LSW) technique, wherein photons are converted into dark photons, which then pass through the wall unimpeded, and then converted back into detectable photons on the far side of the wall. This is a common technique of searching for dark photons independent of the local dark photon abundance and has a natural signal/noise veto by turning the input light source on and off~\cite{jaeckel2010low}. The second method is less common in dark photon detection, but was performed with a superconducting qubit~\cite{Akash2020} and has been proposed in other experiments~\cite{DM_Radio,masha,gelmini2020probing,mina}. This method gains great sensitivity in $\chi$ far below that bound at the expense of a more constrained mass range. Because the interesting masses for the dark photon range from 10 feV to 1 TeV~\cite{darkphotonreview}, there is plenty of room for these experiments to make valuable exclusions or discovery.

Haloscopes originally designed for axion detection~\cite{admx2018,ADMXnov2019,ADMXsidecar,rbf,uf,CAPP,CAPP3,CAPP2,Kelly2020,Kelly2018,HAYSTAC} are also sensitive to dark photons interacting via the Lagrangian in Equation \ref{eq:lagrangian}~\cite{arias2012wispy}. Unlike axion-photon coupling, dark photons have no dependence on a magnetic field, thereby simplifying experiments greatly from haloscope axion searches.
Experimentally installing a magnet is nontrivial, and so as a first test several experiments have carried out dark photon experiments~\cite{Akash2020, phipps2020exclusion}.
A dark photon signal from a relic abundance is resonantly enhanced in the same way as an axion signal, which manifests inside the haloscope as a power fluctuation. Although the conversion of an oscillating dark-photon field to a small oscillating electric field does not require a magnetic field, it is not harmed by its presence, allowing axion haloscopes to simultaneously search for both axions and dark photons. In the haloscope mass region of 1-40 \SI{}{\micro\electronvolt}, the dark photon's kinetic mixing strength is bounded from above by experiments and observations from both Coulomb's law measurements of the photon mass and distortions in the cosmic microwave background~\cite{Coulombsource,CMBsource,arias2012wispy}. 

\section{\label{sec:analysis}Scaling Haloscope Exclusions for Dark Photon Searches}
\begin{table}
\centering
\caption{Values used in Equation \ref{eq:scaling_factor} to calculate the dark photon exclusions from the reported axion exclusions values in the cited papers. Data was lifted from the reported exclusions using WebPlotDigitizer~\cite{webplotdigitizer}.}
\label{tab:values}
\begin{tabular*}{\linewidth}{l @{\extracolsep{\fill}}cllc}
\hline
\hline
        & $m_a$ ($\mu$eV) & $B$ (T) & $\rho_a$ $\left(\frac{\text{GeV}}{\text{cm}^3}\right)$ & Publication \\ \hline
ADMX    & 1.90-3.69       & 7.9     & 0.45                  &~\cite{admx2018} 2018    \\
        & 2.67-3.32       & 7.6     & 0.45                  &~\cite{ADMXnov2019} 2020 \\
        & 17.38-17.57     & 3.11    & 0.45                  &~\cite{ADMXsidecar} 2018 \\
        & 21.03-23.98     & 2.55    & 0.45                  & "                       \\
        & 29.67-29.79     & 3.11    & 0.45                  & "                       \\
RBF     & 4.35-5.60       & 5.8     & 0.3                   &~\cite{rbf} 1989         \\
        & 5.62-7.97       & 7.5     & 0.3                   & "                       \\
        & 8.00-9.97       & 7.0     & 0.3                   & "                       \\
        & 11.13-11.56     & 5.8     & 0.3                   & "                       \\
        & 11.59-16.13     & 7.5     & 0.3                   & "                       \\
UF      & 5.46-7.60       & 7.6     & 0.2805                &~\cite{uf} 1990          \\
CAPP    & 6.62-6.82       & 7.3     & 0.45                  &~\cite{CAPP} 2020        \\
        & 10.16-10.34     & 7.9     & 0.45                  &~\cite{CAPP3} 2021       \\
        & 10.34-11.37     & 7.2     & 0.45                  & "                       \\
        & 12.99-13.88     & 7.8     & 0.45                  &~\cite{CAPP2} 2020       \\
HAYSTAC & 16.96-17.28     & 8.0     & 0.45                  &~\cite{Kelly2020} 2021   \\
        & 23.15-24.0      & 8.0     & 0.45                  &~\cite{Kelly2018} 2018   \\
\hline
\hline
\end{tabular*}
\end{table}

The signal coming out of a haloscope is very narrow in mass, and therefore requires the cavity frequency to be on resonance with the mass being searched in order to ensure maximum signal to noise efficiency.
The on-resonance axion conversion power in the haloscope is given as~\cite{NIM_HAYSTAC}
\begin{equation}
    \mathcal{P} = \left(\frac{g_{a\gamma\gamma}}{m_a}\right)^2 m_a \rho_a B_0^2 V C_{mnl} Q_L \frac{\beta}{1+\beta}
    \label{eq:axion_power}
\end{equation}
Here, $g_{a\gamma\gamma}$ is the axion-photon coupling constant, $m_a$ is the mass of the axion, $\rho_a$ is the local density of the axion field, $B_0$ is the magnetic field~\footnote{Haloscope papers tend to put their magnetic fields in units of Tesla, but Equation \ref{eq:axion_power} requires that the magnetic field be in units of \SI{}{\giga \electronvolt \squared}. Tesla can be converted to square Joules by multiplying by a factor of $1 = \hbar c^2 e/\hbar c^2 e$ where the numerator is in SI units and the denominator is in natural units, i.e. $\hbar=c=1$ and $e=\sqrt{4 \pi \alpha}$. This gives us the magnetic field in units of energy squared, which can then be easily converted to \SI{}{\giga \electronvolt \squared}.}, $V$ is the volume of the cavity, 
$C_{mnl}$ is the mode form factor chosen such that the electric field is aligned with the uniform external magnetic field (required by the $\textbf{E}\cdot\textbf{B}$ interaction in the axion's Primakoff effect~\cite{primakoff}), $Q_L$ is the loaded quality factor of the cavity, and $\beta$ is the cavity coupling.

The signal power in a haloscope from dark photon conversion follows analogously. The power extracted from a haloscope due to dark photon conversion on resonance, as derived in~\cite{arias2012wispy} is
\begin{equation}
    \label{eq:dark_photon_power}
    \mathcal{P} = \chi^2 m_{\gamma'} \rho_{\gamma'} V \mathcal{G} Q_L \frac{\beta}{1+\beta}
\end{equation}
where $\rho_{\gamma'}$ is the local density of the dark photon dark matter, and $\mathcal{G}$ is the form factor of the cavity with respect to the dark photon field, where the component of the dark photon vector field parallel to the electric field takes the place of the B-field.

The dark photon form factor $\mathcal{G}$ is related to the axion form factor $C_{mnl}$ by $\mathcal{G} = C_{mnl} \cos^2{\theta}$~\cite{arias2012wispy} where $\theta$ is the angle between the axis of the cavity and the polarization of the dark photon vector field. Note that, because the dark photon is massive, its wavevector and polarization are independent. This means that a uniform polarization does not necessarily imply a coherent field, and therefore the signal is not enhanced in a polarized field; rather, it may in fact lower the sensitivity of the haloscope.

It can be seen that the axion and the dark photon on-resonance signal power expressions have the same linear dependence on the energy density of the dark matter candidate. The weighting procedure in~\cite{analysisprocedure} also holds for dark photons as the dark photon signal power differs from the axion signal power only by a scaling factor, and the virialized lineshape is assumed to be the same. This means that the post-processed data can be simply scaled into the dark photon parameter space. By setting Equations \ref{eq:axion_power} and \ref{eq:dark_photon_power} equal and using $m_{\gamma'}=m_a$, it becomes clear that the conversion on resonance between the axion coupling constant $g_{a\gamma\gamma}$ and the dark photon kinetic mixing coefficient $\chi$ is
\begin{equation}
    \label{eq:scaling_factor}
    \frac{\chi}{g_{a\gamma\gamma}}=\frac{B_0}{m_{\gamma'} \haschanged{\sqrt{\langle\cos^2{\theta}\rangle}}} \sqrt{\frac{\rho_a}{\rho_{\gamma'}}}
\end{equation}
\haschanged{where $\langle\cos^2{\theta}\rangle$ comes from averaging over the possible dark photon polarizations during the measurement integration time.} In the case that the dark photon field does not have a uniform polarization, or for a sufficiently rapidly varying polarization, the cavity is still only sensitive to the polarization component along the axis of the cavity; there are two other orthogonal polarizations, so the haloscope only measures 1/3 of the dark photons. Therefore the form factor can be averaged over the unit sphere to get $\mathcal{G} = C_{mnl}/3$. For a uniform fixed polarization, however, the dot product between the axis of the cavity and the polarization can vary, which requires a probabilistic analysis. \haschanged{Here we take the simplest possible approach to preserve the originally reported axion confidence limits, which are 90\% for ADMX, CAPP, and HAYSTAC~\cite{admx2018, ADMXnov2019, ADMXsidecar, CAPP, CAPP2, CAPP3, Kelly2018, Kelly2020}, and 95\% for RBF and UF experiments~\cite{rbf, uf}.}

\haschanged{
For a spherical symmetry where the axis of the cavity in the haloscope is aligned with the z-axis, $|\cos\theta|$ of the dark photon polarization is uniformly distributed over $[0,1]$. Here we assume the power fluctuations in the haloscope cavity are Gaussian distributed, with a mean value that is suppressed by the square of $|\cos\theta|$, as in Equation \ref{eq:dark_photon_power} from the factor of $\mathcal{G} = C_{mnl} \cos^2{\theta}$.
To make the calculations easier, we use the square of our uniform random variable, $\cos^2\theta$, which has a PDF of $(2|\cos\theta|)^{-1}$, and normalize our power by subtracting the measured power from it and dividing by the standard deviation, $X=(P - p) / \sigma_P$, so that $X$ is a normal random variable. 
Therefore, our joint PDF between the two random variables is found by multiplying the PDFs of the two random variables. To find the confidence limit for a given value of $x_0$, we integrate this joint PDF over $\cos^2{\theta}$ from $0$ to $1$, then integrate over $x$ from $-\infty$ to $0$ because by definition $X$ has the value of 0:
\begin{equation}
    \label{eq:joint_pdf}
    \small
    \text{CL}=
    \int_{-\infty}^0\int_0^1
    \frac{1}{2\cos\theta}\frac{1}{\sqrt{2\pi}}e^{-\frac{1}{2}\left(x-x_0\cos^2{\theta}\right)^2}
    d\cos^2{\theta} dx.
\end{equation}
Our desired value of the time averaged $\langle\cos^2{\theta}\rangle_T$ is then found by numerically solving for the value of $x_0$ and dividing it from the known value for the standard Gaussian, for example 1.645 for 95\% and 1.272 for 90\%. For an instantaneous measurement, this results in $\cos^2{\theta}$ of 0.076(0.024) for a CL of 90\%(95\%).
}

\haschanged{
For measurements shorter than an hour per run, the effect of the rotation of the earth plays a role only up to a factor of $\sqrt{2}$. For longer measurements, however, the rotation of the earth needs to be taken into account \cite{caputo2021dark} specifically when calculating the PDF of $\cos^2{\theta}$, keeping the rest of the analysis the same. For the purposes of this manuscript, this timing analysis only applies on the deep spike on the CAPP exclusion around \SI{10}{\micro\electronvolt}, the run for which was 15 hours long. For that data, $\cos^2{\theta}$ was found to be 0.29.
}

\haschanged{These calculations along with} Equation \ref{eq:scaling_factor} \haschanged{were} used to scale published axion exclusions using the published experimental parameters. These parameters are shown in Table \ref{tab:values}. For data where different mass regions used different values of the magnetic field, exclusions were calculated separately for each region and then stitched back together.

\section{\label{sec:results}Results}
\begin{figure}
    \centering
    \includegraphics[width=\linewidth]{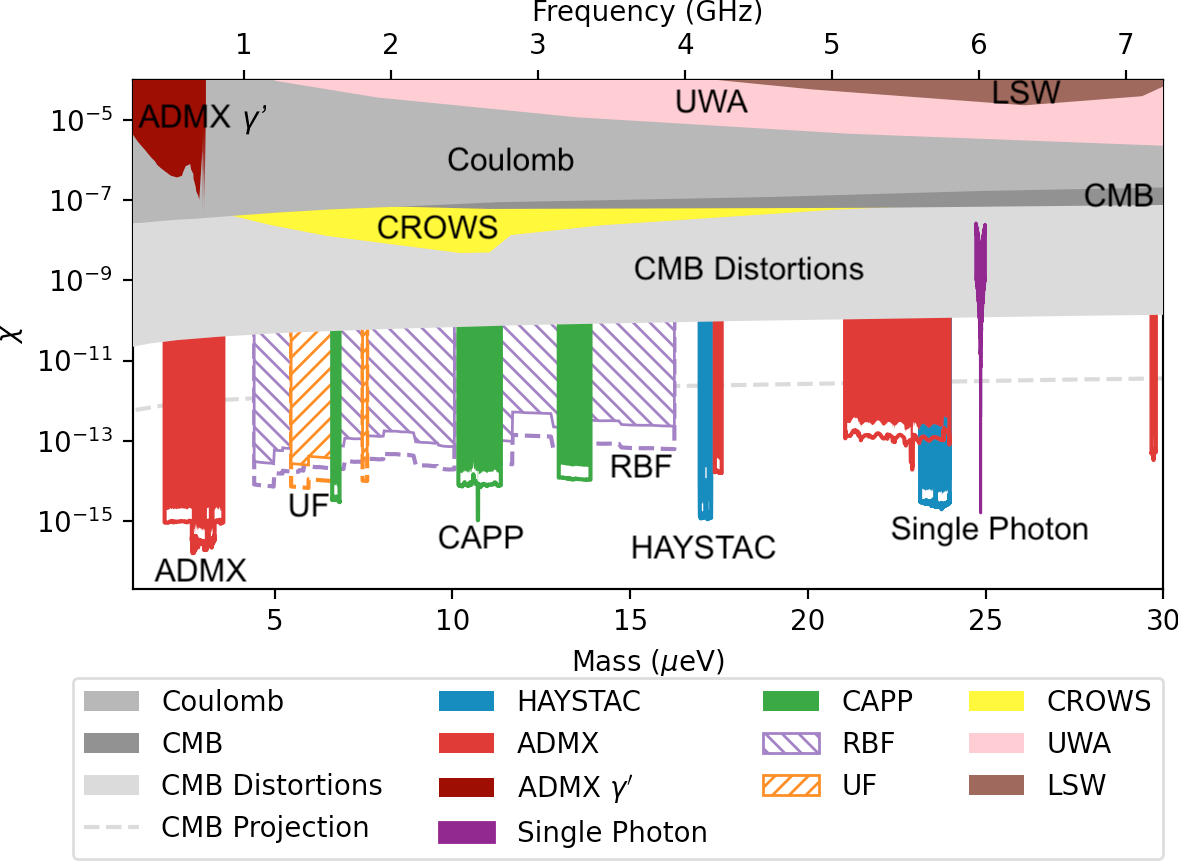}
    \caption{Dark photon exclusions in the haloscope mass range vs the dimensionless kinetic mixing coefficient $\chi$. Haloscope experiments shown are HAYSTAC~\cite{HAYSTAC,Kelly2018,Kelly2020}, ADMX~\cite{admx2018, ADMXnov2019,ADMXsidecar}, Brookhaven~\cite{rbf}, the University of Florida~\cite{uf}, and CAPP~\cite{CAPP,CAPP2,CAPP3}. Outlines for the haloscopes are the exclusions in the case that dark photons have random polarizations; filled in patches are the \haschanged{slowly varying polarized case with confidence limits calculated to match the recorded axion CLs}. Experiments using magnetic veto are hashed out due to unknown gaps in the data. Also shown are ADMX's dark photon experiment~\cite{ADMX_2cavities}, single photon detection~\cite{Akash2020}, LSW~\cite{LSW}, CROWS~\cite{CROWS}, and UWA~\cite{UWA}. Exclusions made from calculations on cosmic microwave background measurements~\cite{CMBsource}, projections of future CMB experiments~\cite{CMB_distortions_update}, and Coulomb's law~\cite{Coulombsource} are also shown. Our sensitivity limits presented here \haschanged{as well as the code to calculate $\cos^2{\theta}$} are publicly available here~\cite{data}.}
    \label{fig:haloscope_parameter_space}
\end{figure}

Figure \ref{fig:haloscope_parameter_space} presents dark photon exclusions in the 1-30~\SI{}{\micro \electronvolt} regime to which haloscopes are sensitive, \haschanged{normalized to a} dark matter density of \SI{0.45}{\giga\electronvolt\per\centi\meter\cubed}~\cite{read2014local}. The haloscope exclusions were calculated using the magnetic fields and dark matter densities recorded in those references for each measured region, tabulated in Table~\ref{tab:values}. The shaded haloscope exclusions represent the conservative case where the dark photon field has a single polarization almost perpendicular to the axis of the haloscope, while the outlined haloscope exclusions represent the case that the polarization of dark photons is random. Experiments that used magnetic veto to eliminate potential axion signals, wherein the magnetic field is turned off when a signal is found in order to eliminate it as an axion candidate, are hashed out but still reported; the eliminated candidates could have been the dark photon signal independent of a magnetic field. The haloscope exclusions are shown alongside dedicated dark photon experiments, where the ADMX dark photon exclusion is shown in dark red~\cite{ADMX_2cavities}, LSW is in brown~\cite{LSW}, UWA is in pink~\cite{UWA}, CROWS is in yellow~\cite{CROWS}, and single photon detection is in purple~\cite{Akash2020}. The remaining regions in various shades of grey are exclusions calculated from the cosmic microwave background~\cite{CMBsource} and Coulomb's law measurements~\cite{Coulombsource} where the lightest grey region covers parameter space for which the dark photon would not be able to comprise all of the dark matter~\cite{CMB_distortions_update}. Finally, the grey dashed line is the projected sensitivity of future CMB experiments in this mass range~\cite{CMB_distortions_update}.

In contrast to the technique highlighted in this manuscript, the experiments in the ADMX dark photon experiment~\cite{ADMX_2cavities} and the LSW experiment~\cite{LSW} use the theory described in~\cite{jaeckel2008cavity}. In this framework, a SM photon kinetically mixes into a dark photon in an emitter cavity, passes through the cavity walls, and reconverts into a SM photon in the detector cavity. This technique does not rely on the relic abundance of dark photons, but rather on two kinetic mixing interactions. However, if dark photons can be measured from a relic abundance, a single-cavity detector is significantly more sensitive in $\chi$, as it only depends on a single kinetic mixing interaction. For comparison, the probability that a dark photon is detected in a two-cavity detector is equal to the ratio of the power measured in the detector cavity over the power emitted into the emitter cavity~\cite{jaeckel2008cavity}
\begin{equation}
    \frac{\mathcal{P}_{\text{detector}}}{\mathcal{P}_{\text{emitter}}} = \chi^4 \left(Q_L (1+\beta)\right)^2 \left(\frac{m_{\gamma'}}{\omega_0}\right)^8 |\mathcal{G}|^2
\end{equation}
\haschanged{for two identical cavities}. The advantages of this technique are that it does not require that dark photons be the dark matter and that it is a broadband search which allows for improved scanning across a larger range of frequencies than a single cavity experiment. When we take the approximation that dark photons dominate the composition of the dark matter, however, a single detection will be more sensitive than a production-detection pair.

Haloscope searches cover a relatively narrow frequency window but are extremely sensitive within that narrow range, enabling them to make exclusions in yet unexplored regions of parameter space. As efforts to improve the frequency scan rate of haloscope experiments continue, they will likely be able to make further contributions to the dark photon search in the future.

\section{\label{sec:conclusion}Conclusions}
Haloscopes are sensitive to and competitive in dark photon parameter space in the mass region of 1-40~\SI{}{\micro \electronvolt}, setting limits on $\chi$ as low as $10^{-16}$. Both the conservative (uniformly polarized) and the non-conservative cases for the dark photon field were considered. The method outlined in this work for using a single cavity haloscope as a dark photon detector may be applicable to any haloscope that employs a similar analysis procedure. Our calculated exclusion limits in Figure \ref{fig:haloscope_parameter_space} \haschanged{as well as code to calculate $\cos^2{\theta}$ for each limit} are available at \cite{data}. Future work will \haschanged{include potential improvements in signal strength by including off-resonant contributions \cite{chaudhuri2019optimal}. The dark photon limits in the polarized case can be enhanced by tailoring the method of conversion to each haloscope experiment's analysis method, particularly by exploiting the overlap between subsequent measurements.} Similarly to axion searches, this technique will be greatly enhanced by single photon detection \cite{lamoreaux2013analysis,Akash2020}, as well as future improvements in squeezed-state receivers \cite{malnou2019squeezed, CEASEFIRE}.

\section{Acknowledgements}
We would like to thank Kelly Backes, Danielle Speller, Karl van Bibber, Yonathan Kahn, Masha Baryakhtar, and our anonymous reviewer for fruitful discussions and feedback. We would also like to acknowledge Annie Giman in preparing the manuscript. SG is supported in part by the National Science Foundation under grant number DMR-1747426; SG, MJ, and RM are supported in part by Department of Energy under grant number DE-AC02-07CH11359. A. L. is supported in part by the National Science Foundation under Grant No. PHY-1914199.

\bibliographystyle{ieeetr}
\bibliography{citations}

\end{document}